# Dynamics of topological light states in spiraling structures

Yaroslav V. Kartashov,[1,2,*] Victor A. Vysloukh,[1] and Lluis Torner[1]

[1]ICFO-Institut de Ciencies Fotoniques, and Universitat Politecnica de Catalunya, 08860 Castelldefels (Barcelona), Spain
[2]Institute of Spectroscopy, Russian Academy of Sciences, Troitsk, Moscow Region, 142190, Russia

We expose a mechanism for the dynamical generation and control of light states with diverse topologies in spiraling guiding structures. Specifically, we show that spiraling shallow refractive index landscapes induce coupling and periodic energy exchange between states with different topological charges. Such a resonant effect enables excitation of optical vortices by vortex-free inputs and allows the output topological charge of the beam to be controlled. The presence of nonlinearity results in a strong asymmetrization of the resonant curves and a shift of the resonant frequencies. Resonant vortex dynamic generation, including revivals, is shown to be possible not only in waveguides mediated by total internal reflection but also in Bragg-guiding hollow-core geometries.

The global and local topology of the wavefront of a light field is a fundamental property of the corresponding beam, with profound implications for its evolution. The simplest examples are topological phase singularities nested in a smooth wavefront, which appear as vortices around which the light energy flow circulates [1]. Vortices can follow complex trajectories and even form knots [2,3]. Light beams carrying vortices are employed in numerous applications (for a review see, e.g., Ref. [4]).

Vortices can be nested in light beams in a variety of ways [5]. However, their dynamical evolution depends critically on the properties of the medium. Thus, vortices may be highly sensitive to non-uniformities of the refractive index in the propagation path. Here we show that suitable spiraling guiding structures offer new tools for controllable dynamical transformation of the vortex content of propagating beams that are not accessible in uniform materials. The key insight put forward is based on the dynamics of topological states of light in modulated guiding structures where the orbital angular momentum of light, and hence the field topology, is not conserved.

Longitudinal refractive index modulations may lead to a variety of resonant effects (for a recent survey see [6]). In particular, revivals, akin to stimulated transitions in two-level quantum systems [7], were implemented in long-periodic gratings created in optical fibers [8,9]. Stimulated conversions of one-dimensional guided modes were demonstrated in shallow waveguides and periodic lattices [10,11]. Of special interest is the realization of stimulated transition between states having different topologies, for example between vortices with different topological charges. Resonant excitation of vortices has been studied in optical fibers wrapped in a coiled wire of a constant pitch [12], fibers with tilted gratings written by interfering UV beams [13], stressed [14] or helical fibers drawn from a preform with an off-centered core [15,16], and in chiral fiber gratings [17,18]. Resonances of light fields with nonzero orbital angular momentum have been observed recently in solid-core helically twisted photonic-crystal fibers [19,20]. However, dynamic excitation of vortex states has not been analyzed in shallow guiding structures using the full model accounting for energy leakage through radiation and indirect parametric-mode conversions. Resonant excitation of vortices has not been addressed in Bragg guides, where modes are supported by the Bragg scattering rather than by total internal reflection. Here we show that vortices with various topological charges can be excited by vortex-free modes in spiraling Bragg-type or ordinary waveguides at specific spiraling rates. We perform a broadband scanning of spiraling frequencies and show how the output wavefront topology can be varied by shitting the detuning around the resonances.

We describe the propagation of light beams along the $\xi$-axis of a medium with shallow dynamical modulation of the refractive index by Schrödinger equation for the normalized light field amplitude $q$:

$$i\frac{\partial q}{\partial \xi} = -\frac{1}{2}\left(\frac{\partial^2 q}{\partial \eta^2} + \frac{\partial^2 q}{\partial \zeta^2}\right) + \sigma q|q|^2 - pR(\eta,\zeta,\xi)q. \qquad (1)$$

Here the transverse coordinates $\eta,\zeta$ are normalized to the characteristic transverse scale $r_0$; the longitudinal coordinate $\xi$ is normalized to $kr_0^2$, where $k=2\pi n/\lambda$ is the wavenumber; the nonlinear parameter $\sigma>0$ ($\sigma<0$) corresponds to defocusing (focusing) nonlinearity; the parameter $p=\delta n k^2 r_0^2/n$ is proportional to the real variation of the refractive index $\delta n$; $R$ is the function describing the refractive index profile. For $r_0=10\,\mu\mathrm{m}$ and light beams at the wavelength $\lambda=633\,\mathrm{nm}$ propagating in fused silica [21], a waveguide depth $p=10$ corresponds to a refractive index contrast $\delta n \sim 7\times 10^{-4}$, the distance $\xi=1$ corresponds to a propagation distance of $1.4\,\mathrm{mm}$ and $|q|^2=1$ corresponds to peak intensity $I\sim 2.6\times 10^{15}\,\mathrm{W/m^2}$ assuming a nonlinear coefficient $n_2=2.7\times 10^{-20}\,\mathrm{m^2/W}$.

First, we consider resonant excitation of vortex modes in a Gaussian waveguide. Examples of the eigenmodes supported by a non-spiraling waveguide $R = \exp(-r^2)$ in a linear medium ($\sigma = 0$), are shown in Fig. 1(a). For $p = 10$ there exist three eigenmodes $e_m(r,\phi,\xi) = w_m(r)e^{im\phi + ib_m\xi}$ [here $r = (\eta^2 + \zeta^2)^{1/2}$ and $\phi$ are the radius and azimuthal angle], without radial nodes, with propagation constants $b_0 \approx 6.042$, $b_1 \approx 2.685$, $b_2 \approx 0.141$, and topological charges $m = 0,1,2$. In a non-spiraling waveguide the evolution of any combination $q(\eta,\zeta,\xi) = \sum_m C_m e_m$ of these modes results in the formation of a complex dynamically varying pattern (for example, the combination of $e_1$ and $e_2$ modes yields a beam with one centered and one off-center phase dislocation that rotates upon propagation), but the mode energy weights $\nu_m(\xi) = |C_m|^2$ remain constant.

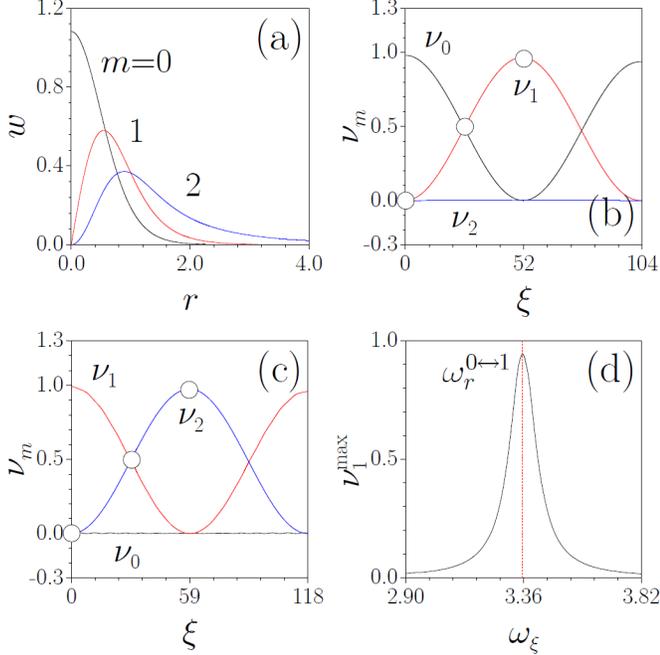

Fig. 1. (a) Eigenmodes of a Gaussian waveguide with different topological charges, for $p = 10$. The mode weights $\nu_m$ with $m = 0,1,2$ versus $\xi$ for $\omega_\xi = \omega_r^{0 \leftrightarrow 1}$ (b) and $\omega_\xi = \omega_r^{1 \leftrightarrow 2}$ (c). In (b) only the $m = 0$ mode was launched into the waveguide at $\xi = 0$, while in (c) only the $m = 1$ mode was launched. Circles in (b) and (c) correspond to distributions from Figs. 2(a) and 2(b), respectively. (d) The dependence of the maximal weight of the $m = 1$ mode on $\omega_\xi$. In all cases $\mu = 0.01$.

The light evolution changes drastically if the refractive index experiences periodic variation along $\xi$, because it introduces coupling and thus energy exchange between the modes. If the waveguide remains symmetric only coupling between modes of the same parity occurs, thus such type of modulation does not lead to coupling of modes with different topological charges. In contrast, such coupling can be realized in spiraling waveguides with $R(\eta,\zeta,\xi) = \exp\{-[\eta + \mu\sin(\omega_\xi\xi)]^2 - [\zeta - \mu\cos(\omega_\xi\xi)]^2\}$, where $\mu \ll 1$ is the radius and $\omega_\xi$ is the spatial spiraling frequency. This coupling is resonant, and the efficiency of energy exchange between two modes is maximal at the frequencies corresponding to the difference of their propagation constants, i.e. at $\omega_r^{0 \leftrightarrow 1} = b_0 - b_1$ and $\omega_r^{1 \leftrightarrow 2} = b_1 - b_2$.

For small radius of spiraling $\mu \ll 1$ we look for distance-dependent superposition of the fundamental ($m = 0$) and vortex ($m \neq 0$) modes: $q(\eta,\zeta,\xi) = C_0(\xi)w_0(r)e^{ib_0\xi} + C_m(\xi)w_m(r)e^{im\phi}e^{ib_m\xi}$. We expand the refractive index profile $R$ in series with respect to small displacement of the waveguide center and keep only the first-order terms in the expansion that are proportional to $\mu$. Application of the technique of resonant mode coupling [10] yields the system of equations $idC_0/d\xi = \mu p \mathcal{I}_r \mathcal{I}_\phi C_m e^{i(b_m - b_0 + \omega_\xi)\xi}$, $idC_m/d\xi = \mu p \mathcal{I}_r \mathcal{I}_\phi^* C_0 e^{i(b_0 - b_m - \omega_\xi)\xi}$ in which $\mathcal{I}_r = \int_0^\infty w_0 w_m e^{-r^2} r^2 dr$, $\mathcal{I}_\phi = \int_0^{2\pi} e^{i(m-1)\phi} d\phi$ are the exchange integrals. An important conclusion is that within this approximation, energy transfer and vortex formation are possible at resonant spiraling frequency $\omega_\xi = b_0 - b_m$ only when $m = 1$ (or $m = -1$ if the waveguide spirals in the opposite direction), while the distance $L = \pi/(\mu p \mathcal{I}_r |\mathcal{I}_\phi|)$ of vortex formation is defined by the overlap integrals $\mathcal{I}_r, \mathcal{I}_\phi$.

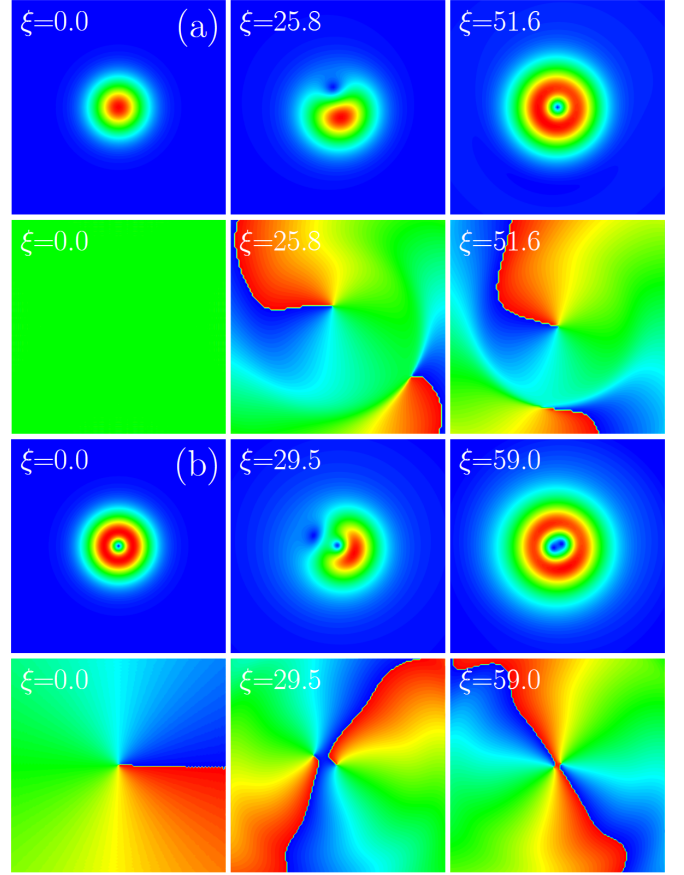

Fig. 2. Intensity and phase distributions at different distances, illustrating the dynamics of resonant conversion of (a) the $m = 0$ mode into the $m = 1$ mode for $\omega_\xi = \omega_r^{0 \leftrightarrow 1}$ and of (b) the $m = 1$ mode into the $m = 2$ mode for $\omega_\xi = \omega_r^{1 \leftrightarrow 2}$. In all cases $\mu = 0.01$.

Numerical simulations based on Eq. (1) confirm the coupled-mode predictions. Using weight coefficients defined as $\nu_m(\xi) = \left|\iint_{-\infty}^{+\infty} e_m^*(\eta,\zeta,\xi) q(\eta,\zeta,\xi)\, d\eta d\zeta\right|$, we illustrate in Fig. 1(b) the process of energy exchange between the $m = 0$ and $m = 1$ states. The spiraling of the waveguide causes slow periodic transfer of power between fundamental and $m = 1$ modes, while the $m = 2$ mode remains unexcited. In such an adiabatic regime the efficiency of vortex excitation is high, as nearly all input power may be transferred to the vortex mode when the resonance condition is fulfilled. The sign of the topological charge of the vortex is dictated by the direction of spiral-ing. The conversion process is illustrated in Fig. 2(a). The phase singularity always appears in the low-intensity region and then gradually moves to the axis of the beam. Spiraling of the waveguide results in leakage losses. For $\mu \sim 0.01$, radiative losses over the distance $\xi_{tr}=104$ (one period of complete energy exchange) do not exceed 5%, but they grow when $\omega_\xi$ increases. Resonance between modes with $m=1$ and $m=2$ [Fig. 1(c)] allows further increasing of the topological charge. As in the case of the $0 \leftrightarrow 1$ transition, for the $1 \leftrightarrow 2$ transition the second phase dislocation comes from the field periphery (conceptually, from the transverse infinity) and then fuses with the central one, resulting in the doubling of the topological charge at a certain distance [Fig. 2(b)].

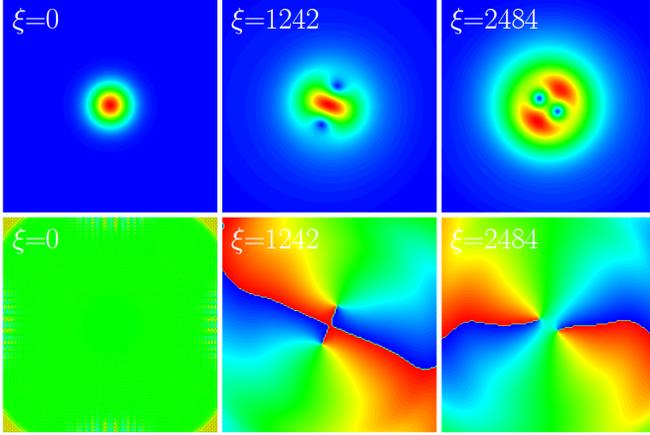

Fig. 3. Intensity and phase distributions at different propagation lengths, illustrating the dynamics of conversion of the $m=0$ mode into the $m=2$ mode at $\omega_\xi = \omega_r^{0\leftrightarrow 2}/2$ and $\mu=0.006$.

The direct transitions $0 \leftrightarrow 1$ and $1 \leftrightarrow 2$ lead to almost complete power transfer into vortex modes with higher topological charges, while the direct transition $0 \leftrightarrow 2$ does not occur thus the vortex mode with $m=2$ is not excited at the spiraling frequency $\omega_r^{0\leftrightarrow 2}=b_0-b_2$. It is possible to excite the $m=2$ vortex by using the cascading transformation $0 \to 1 \to 2$, which implies changing the spiraling frequency $\omega_\xi$ from $\omega_r^{0\leftrightarrow 1}$ to $\omega_r^{1\leftrightarrow 2}$ at the distance where almost all power is transferred from the input $m=0$ mode into the $m=1$ mode. However, the $m=2$ mode is also excited by the $m=0$ mode due to parametric resonance at the fractional frequency $\omega_r^{0\leftrightarrow 2}/2$. In order to describe such parametric resonances one has to take into account the second-order terms $\sim \mu^2$ in the expansion series of the refractive index profile for small displacements. In this case the coupled mode approach yields $idC_0/d\xi = -(\mu^2 p/2)\mathcal{I}_r \mathcal{I}_\phi C_m e^{i(b_m-b_0+2\omega_\xi)\xi}$ and $idC_m/d\xi = -(\mu^2 p/2)\mathcal{I}_r \mathcal{I}_\phi^* C_0 e^{i(b_0-b_m-2\omega_\xi)\xi}$, where the exchange integrals are given by $\mathcal{I}_r = \int_0^\infty w_0 w_m e^{-r^2} r^3 dr$ and $\mathcal{I}_\phi = \int_0^{2\pi} e^{i(m-2)\phi} d\phi$. The strength of this interaction is proportional to $\mu^2$, indicating that considerable distances are required for power transfer. The dynamics in the case of parametric resonance $0 \leftrightarrow 2$ are shown in Fig. 3. Now both phase dislocations, which appear at opposite sides of the beam, gradually move toward its axis and nearly fuse there. The complete fusion does not occur because at the parametric resonance a small ($<5\%$) fraction of the power remains in the $m=0$ mode.

The resonant character of the phenomenon discussed here is confirmed by Fig. 1(d), which shows the maximal weight $\nu_1^{max}$ of the $m=1$ mode for the $0 \leftrightarrow 1$ transition versus spiraling frequency. The conversion efficiency decreases with the growth of detuning from resonance $\omega_r^{0\leftrightarrow 1}$. The width of the resonance curve is determined by the radius $\mu$ of spiraling - the increase of $\mu$ results in the broadening of all resonances, as shown in Fig. 4(a) (the widths of the curves $\delta\omega^{0\leftrightarrow 1}$, $\delta\omega^{1\leftrightarrow 2}$ were determined at the level where $\nu_m^{max}=1/2$). The resonance $1\leftrightarrow 2$ is slightly narrower than the $0\leftrightarrow 1$ resonance. The parametric resonance $0\leftrightarrow 2$ is much narrower than all direct resonances. The propagation distance at which the complete cycle of energy exchange between $m=0$ and $m=1$ modes occurs is a monotonically decreasing function of $\mu$ [Fig. 4(b)]. Therefore, on the one hand for faster energy transfer large $\mu$ values are favorable, but at the same time increasing $\mu$ stimulates radiative losses.

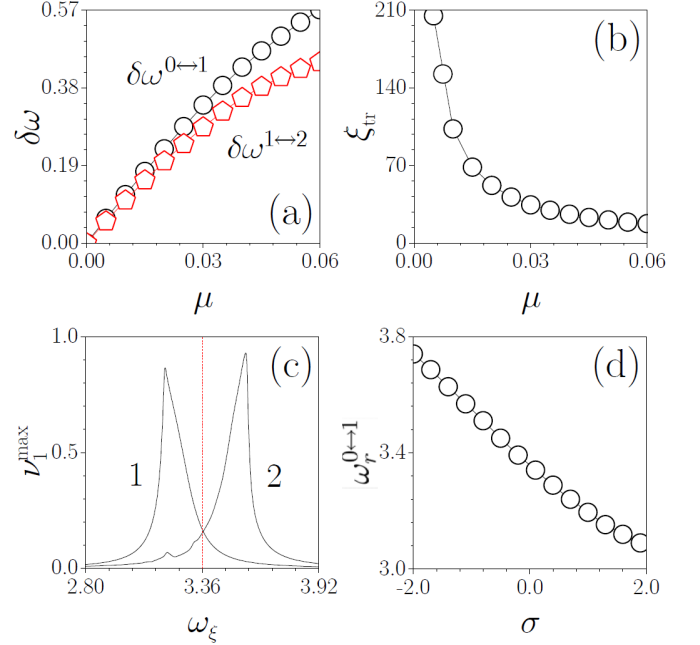

Fig. 4. (a) Width of the resonance for $0 \leftrightarrow 1$ and $1 \leftrightarrow 2$ transitions versus $\mu$ at $\sigma = 0$. (b) Distance of energy exchange for the $0 \leftrightarrow 1$ transition versus $\mu$ at $\sigma = 0$. (c) Maximal weight of the mode with $m=1$ versus $\omega_\xi$ for the $0 \leftrightarrow 1$ transition at $\sigma = +1.1$ (curve 1) and $\sigma = -1.1$ (curve 2) at $\mu = 0.01$. The dashed line indicates the resonance frequency for a linear medium. (d) Resonance frequency $\omega_r^{0\leftrightarrow 1}$ as a function of $\sigma$ for $\mu = 0.01$.

We found that resonant transitions are possible even for relatively high nonlinearities ($|\sigma| \sim 2$) in both focusing and defocusing media. Nonlinearity results in a drastic deformation of the resonance curves - they become notably asymmetric. For defocusing nonlinearities, the low-frequency wing of the resonance experiences steepening, while the high-frequency wing flattens [curve 1 in Fig. 4(c)]. In contrast, in focusing media the opposite situation occurs: the high-frequency wing becomes steeper than the low-frequency one [curve 2 in Fig. 4(c)]. The deformation of the resonance curve becomes more pronounced when $|\sigma|$ grows, while the maximal conversion efficiency decreases with $|\sigma|$ because nonlinearity stimulates radia-tion. Nonlinearity also shifts all resonance frequencies because it generates different shifts of propagation constants of the different modes. In a focusing medium the resonance frequency increases, while in a defocusing medium it decreases [Fig. 4(d)]. In the range of nonlinear coefficients $|\sigma|<2$ considered here azimuthal instability for vortices does not appear at least for $\xi<5\xi_{\rm tr}$ due to the stabilizing action of the waveguide.

Finally, we found that dynamic topology control is possible not only in conventional waveguides but also in Bragg-guiding structures with a hollow core surrounded by concentric rings, which are described by the function $R(r)=0$ for $r<(2k-1)\pi/2\Omega$ and by $R(r)=\cos^2(\Omega r)$ for $r\geq(2k-1)\pi/2\Omega$, where $k\in\mathbb{N}$. The linear eigenmodes of such a structure with topological charges $m=0,1$ are shown in Fig. 5(a) for $p=2.5$, $\Omega=2$, and $k=2$. Notice the presence of long oscillating tails in the mode profiles, which are typical for the gap-type localization. In a non-spiraling waveguide such modes propagate independently, but in the presence of spiraling perturbations (we assume that the center of the waveguide moves on the circle of radius $\mu$ with frequency $\omega_\xi$) stimulated conversion occurs at the resonance frequency $\omega_r^{0\leftrightarrow1}=b_0-b_1$, where the propagation constants are now negative [6]: $b_0\approx-0.409$, $b_1\approx-0.896$. The resonance curve for the $0\leftrightarrow1$ transition is shown in Fig. 5(b), while the dynamics of transition is illustrated in Fig. 6. Phase dislocations appear due to destructive interference and move from the periphery towards the center. Notice the nontrivial phase distribution of the output $m=1$ mode and the fact that it acquires a multi-ring structure despite considerable deformations that occur at intermediate distances. Because the core of the Bragg waveguides is empty, transfer of angular momentum to the beam occurs via the outer rings and the efficiency of this process is lower than in usual waveguides, so that for a fixed $\mu$ the energy exchange distance between modes is larger in hollow-core guides.

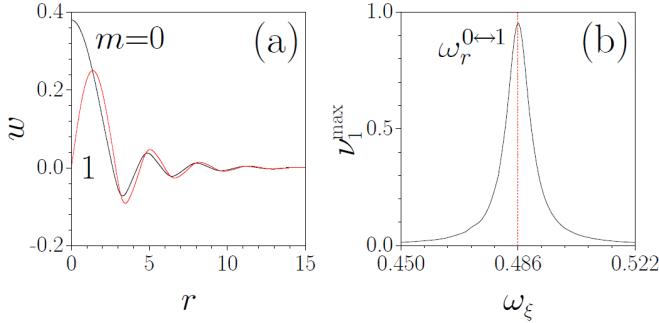

Fig. 5. (a) Eigenmodes of a Bragg waveguide with various charges. (b) Resonance curve for $0\leftrightarrow1$ transition at $\mu=0.01$.

Summarizing, we put forward the possibility of dynamic excitation and control of vortex states in spiraling shallow refractive index landscapes that include radiation and parametric-mode resonances as well as nonlinearity. The dynamic control of the beam topology is accessible not only in total-internal-reflection structures but also in Bragg-guiding geometries.

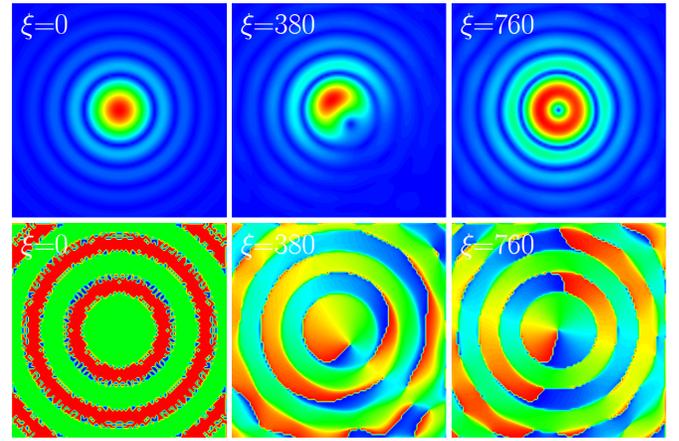

Fig. 6. Intensity and phase distributions at different propagation lengths, illustrating the dynamics of resonant conversion of the $m=0$ mode into the $m=1$ mode for $\omega_\xi=\omega_r^{0\leftrightarrow1}$ and $\mu=0.01$ in a Bragg waveguide.